\documentclass{optica-article}

\journal{opticajournal} 

\articletype{Research Article}

\usepackage{amsmath,mathtools}
\usepackage{bm}

\usepackage{booktabs}
\usepackage{enumitem}

\usepackage{lineno}

\graphicspath{{./}}

\newcommand{\qs}{\mathbf{q}_s}
\newcommand{\qt}{\mathbf{q}_t}
\newcommand{\ps}{\mathbf{p}_s}
\newcommand{\pt}{\mathbf{p}_t}
\newcommand{\nhat}{\hat{\mathbf{n}}}
\newcommand{\shat}{\hat{\mathbf{s}}}
\newcommand{\that}{\hat{\mathbf{t}}}
\newcommand{\dhat}{\hat{\mathbf{d}}}
\newcommand{\dd}{\mathrm{d}}
\newcommand{\R}{\mathbb{R}}
\newcommand{\Sph}{\mathbb{S}}
\newcommand{\thv}{\bm{\theta}}

\begin{document}

\title{Direct Optimization of a 3D Finite-Source Reflector via Neural-Network Parameterization}

\author{Roel Hacking,\authormark{1,*} Lisa Kusch,\authormark{1} Martijn Anthonissen,\authormark{1} and Wilbert IJzerman\authormark{1,2}}

\address{\authormark{1}Eindhoven University of Technology, PO Box 513, 5600 MB Eindhoven, The Netherlands\\
\authormark{2}Signify, High Tech Campus 7, 5656 AE Eindhoven, The Netherlands}

\email{\authormark{*}r.g.j.hacking@tue.nl}

\begin{abstract*}
We present a direct optimization method for three-dimensional freeform reflectors that transform the light of a finite-\'etendue source into a prescribed far-field angular intensity distribution.
The reflector profile is represented by a small neural network (a multilayer perceptron), which is trained end-to-end through a differentiable ray-tracing objective.
We furthermore parameterize the emission directions in gnomonic coordinates, and show how we use this to ensure that every emitted ray intersects the reflector.
At each iteration, the network is converted to a bicubic spline representation for ray-tracing efficiency, and intersections with this smooth surface are solved by a damped Newton solve---with gradients computed via the implicit function theorem.
The traced output distribution is compared with the desired target on a `soft' histogram, under an $H^{-1}$-type spectral weighting that emphasizes long-range transport of flux to improve convergence.
Optimization is performed using a BFGS method with self-scaled Broyden updates and a plateau-perturbation rule to prevent stalling.
The method converges reliably within seconds on a single GPU for all examples tested.
\end{abstract*}

\section{Introduction}\label{sec:intro}

Many optical applications---including illumination, beam shaping, and solar concentration---depend on precise transformation of some source light distribution into some desired target distribution~\cite{WelfordWinston1,WinstonMinanoBeneitez2,Chaves3}.
To produce such transformation, we wish to design optical components, such as reflectors and lenses with freeform surfaces, that effectively realize these desired light transformations.
Most commonly, these components are designed under the simplifying assumption of zero-\'etendue sources, i.e., point or parallel sources.
Under these simpler, idealized assumptions, the reflector design problem can be cast as an optimal transport problem: Glimm and Oliker~\cite{GlimmOliker6} formulated the single-reflector problem as a Monge--Kantorovich mass transfer problem, and Gangbo and Oliker~\cite{GangboOliker7} showed the existence of optimal maps for reflector-type problems.
Romijn et al.~\cite{Romijn8,Romijn9} developed numerical methods for freeform lens and reflector design, under the assumption of point sources and far-field targets, while B\"osel et al.~\cite{BoselGross11,BoselWorkuGross12} contributed ray-mapping techniques for the design of freeform surfaces for collimated beams.
Neural networks have also been used to solve the Monge--Amp\`ere equation---the partial differential equation characterizing these zero-\'etendue formulations---with its transport boundary condition~\cite{Hacking29}, and to directly predict freeform surface topologies that produce a prescribed irradiance distribution~\cite{CerpentierMeuret28}.

In practice, however, most light sources---such as LEDs---have non-zero \textit{\'etendue}.
That is, they have both spatial \textit{and} angular extent.
The resulting target distribution becomes a blurred version of the ideal point or parallel source response~\cite{WeiZhuLiMa5}.
Due to the conservation of \'etendue, this blurring imposes a fundamental limit on the achievable sharpness of any target distribution, and the simpler optimal transport formulation no longer applies---at least not directly.
Several methods have been proposed to account for this finite-\'etendue blurring resulting from real-world light sources.
For example, Fournier et al.~\cite{FournierCasserlyRolland17} construct the reflector as a point-source solution---via the supporting-ellipsoid method of Kochengin and Oliker~\cite{KochenginOliker44}---for a `virtual' target distribution.
They then iteratively adjust this virtual target until a finite-source ray trace reproduces the prescribed target distribution.
Benamou et al.~\cite{Benamou10} later made this search over virtual targets gradient-based, minimizing an entropic optimal-transport error through the same type of point-source parameterization---though at the cost of an entropic transport solve in every loss evaluation, a restriction to convex reflectors, and, so far, a restriction to the two-dimensional setting.

Another, related, strategy comes from signal processing; we treat the finite-source effect as a black-box convolution and use generic iterative deconvolution methods to counteract the smearing effects imposed by the finite source.
For example, we might use additive iterative deconvolution methods---such as van Cittert~\cite{VanCittert18,BurgerVanCittert19}---or multiplicative iterative methods---such as Richardson--Lucy~\cite{Richardson20,Lucy21}.
Kronberg~\cite{Kronberg24} designed an inverse freeform reflector design framework which uses such iterative deconvolution methods to account for scattering-surface effects, which are similar to the blurring effects of finite sources.
The general framework used by Kronberg~\cite{Kronberg24} can also be applied to finite-source effects, as shown for the two-dimensional case in~\cite{Hacking41} and, via a hybrid neural--deconvolution scheme, for the three-dimensional case in~\cite{Hacking42}.
However, such deconvolution methods require solving a zero-\'etendue subproblem in every iteration, which is a one-dimensional ordinary differential equation in the two-dimensional setting~\cite{Hacking41}, but becomes a two-dimensional optimal transport or partial differential equation (PDE) problem in the fully three-dimensional case, unless we assume rotational symmetry.
Thus, the deconvolution route becomes less attractive in the fully three-dimensional setting, due to the high per-iteration cost of solving the zero-\'etendue subproblem.
Moreover, we have no guarantee that the deconvolution will converge to a solution within this setting, and may even diverge for certain problems.

A third line of work applies differentiable ray tracing to optical design problems.
This approach formulates the forward ray-tracing pipeline as a differentiable function, and uses some combination of automatic differentiation and manual adjoint derivation to compute gradients of the optical design objective with respect to the surface parameters.
This then allows us to use first- or second-order optimization methods to optimize the surface parameters directly, though the resulting optimization problem may be highly non-convex.
Much of the research within this differentiable ray tracing and rendering framework has come from the computer graphics community~\cite{Li45,NimierDavid46}.
Here, the goal is generally to optimize a variety of scene parameters---such as geometry, light positions, and material parameters---to match some desired rendered image.
This general approach can also be applied to optical design problems.
For example, de Koning et al.~\cite{DeKoning26} and Heemels et al.~\cite{Heemels25} combined automatic differentiation with B-splines for freeform optical optimization with finite sources, with the former using a custom differentiable ray tracer and the latter using the Mitsuba~3 renderer~\cite{Jakob47}.
Closer to our setting, Sun et al.~\cite{SunDengZhang32} optimized mesh-based reflectors and lenses end to end through differentiable ray tracing, with face-based optimal-transport updates that pull the surface out of poor local optima, and the DiffRayFlow engine~\cite{WangChangWuMaXie33} combines discrete optimal transport with differentiable ray tracing for freeform design.

This paper presents a fully three-dimensional method for finite-source reflector design with a far-field target distribution.
Our main contributions are as follows:
\begin{enumerate}[nosep]
    \item A \textbf{fully three-dimensional formulation} treating the four-dimensional source phase space (2D spatial $\times$ 2D angular) directly, parameterized so that every emitted ray is guaranteed to intersect the reflector.
    \item A \textbf{mesh-free differentiable forward ray-tracing pipeline}: the reflector profile, parameterized by a multilayer perceptron (MLP), is converted to a bicubic spline representation each iteration, rays are intersected with this smooth surface by a damped Newton solve, and gradients are propagated back through the intersection via the implicit function theorem.
    \item An \textbf{$H^{-1}$-weighted soft-histogram loss}: `soft' histograms of the desired target distribution and of the ray-traced output distribution are compared in the frequency domain, in a way that emphasizes long-range transport of flux to improve convergence.
    \item \textbf{Three numerical examples}, every solve completing in 2 to 8 seconds of wall-clock time on a single GPU, demonstrating the method's performance and robustness.
\end{enumerate}

The remainder of the paper is organized as follows.
Section~\ref{sec:formulation} formulates the problem, Section~\ref{sec:method} explains the method, Section~\ref{sec:results} presents some numerical results, Section~\ref{sec:discussion} discusses our findings, and we finally conclude in Section~\ref{sec:conclusion}.

\section{Problem formulation}\label{sec:formulation}

We formulate the reflector design problem in the geometric-optics regime: we model light as rays that propagate in straight lines and reflect off specular surfaces according to the law of reflection.
The source model and the coordinate conventions that we use are introduced in Section~\ref{sec:source}, the way we parameterize the reflector is described in Section~\ref{sec:reflector}, the ray--reflector intersection and the guarantee that every emitted ray hits the reflector are discussed in Section~\ref{sec:coverage}, and the forward ray-tracing routine and optimization objective are presented in Section~\ref{sec:design}.

\subsection{Physical setting and source model}\label{sec:source}

We model the light source and the reflector in a Cartesian coordinate system, with the source plane at $z = 0$.
The reflector is a specular surface above this source plane, and our goal is to design it such that the resulting far-field angular intensity distribution matches some prescribed target distribution.
We follow some of the notation of~\cite{Hacking41,Hacking42}, and an overview of the symbols used in this paper is given in Table~\ref{tab:notation}.

The planar source occupies a compact region $\Omega \subset \R^2$ on the $z = 0$ plane.
Each point $\qs = (q_{s,1}, q_{s,2}) \in \Omega$ emits light into a set of directions $\shat \in \Sph^2_+$, where $\Sph^2_+ = \{\shat \in \R^3 : |\shat| = 1, s_3 > 0\}$ is the open upper unit hemisphere.
We represent these emission directions by their gnomonic coordinates
\begin{equation}\label{eq:gnomonic}
\ps = G(\shat) := \frac{(s_1, s_2)}{s_3} \in \R^2,
\end{equation}
with the inverse
\begin{equation}\label{eq:gnomonic_inv}
G^{-1}(\ps) = \frac{(p_{s,1},\, p_{s,2},\, 1)}{\sqrt{1 + |\ps|^2}}.
\end{equation}
Geometrically, $\ps$ is the central projection of $\shat$ from the origin onto the tangent plane $z = 1$, such that $|\ps| = \tan\alpha$ where $\alpha$ is the polar angle between $\shat$ and the $z$-axis.

The set of admissible emission directions, expressed in gnomonic coordinates, forms the angular source domain $A \subset \R^2$.
The domain~$A$ represents the angular extent of the source: a small $A$ corresponds to a nearly collimated beam, whereas a large $A$ corresponds to a wide-angle emitter.
The \'etendue of the source is determined jointly by $\Omega$ and~$A$.
The full source phase space is $\mathcal{S} = \Omega \times A \subset \R^4$, and the source radiance is described by a normalized distribution function $f : \mathcal{S} \to \R_{\geq 0}$ satisfying $\int_{\mathcal{S}} f(\qs, \ps)\, \dd\qs\, \dd\ps = 1$.

In related finite-source formulations~\cite{Hacking42}, source emission directions were parameterized by stereographic projections from the south pole.
Here we adopt gnomonic coordinates because the ray--reflector intersection equation~\eqref{eq:shadow} becomes \textit{affine} in the emission angular coordinate~$\ps$, which considerably simplifies the ray-coverage analysis of Section~\ref{sec:coverage} and the boundary conditions.
Stereographic and gnomonic coordinates are related by a smooth, invertible change of coordinates on $\Sph^2_+$; they therefore represent the same physical ray directions.
For the \textit{reflected} (output) directions, we retain the north-pole stereographic projection.

\subsection{Reflector surface}\label{sec:reflector}

A reflector over~$\Omega$ is given by the parametric map (cf.~\cite{Hacking41,Hacking42})
\begin{equation}\label{eq:reflector}
\mathbf{r}(\qt) = (\qt, 0) + u(\qt)\, \dhat(\qt),
\end{equation}
where $\qt = (q_{t,1}, q_{t,2}) \in \Omega$ is the reflector surface parameter, $u : \Omega \to \R_{>0}$ is the unknown profile function, and $\dhat : \Omega \to \Sph^2_+$ is a prescribed field of unit directions.
The direction field is constructed from its gnomonic representation $\boldsymbol{\beta}(\qt) = (d_1/d_3, d_2/d_3) \in \R^2$ via
\begin{equation}\label{eq:direction}
\dhat(\qt) = G^{-1}(\boldsymbol{\beta}(\qt)) = \frac{(\beta_1,\, \beta_2,\, 1)}{\sqrt{1 + |\boldsymbol{\beta}|^2}}.
\end{equation}
Note that $u$ is \textit{not} a simple height function: the reflector point $\mathbf{r}$ lies at distance $u$ from the source plane \textit{along the direction}~$\dhat$, not vertically.
The vertical component of the reflector point is $u(\qt) \cdot d_3(\qt)$.

Let $[q_{1,\min}, q_{1,\max}] \times [q_{2,\min}, q_{2,\max}]$ and $[p_{1,\min}, p_{1,\max}] \times [p_{2,\min}, p_{2,\max}]$ be the axis-aligned bounding boxes of $\Omega$ and~$A$; we define the direction field via the componentwise affine map
\begin{equation}\label{eq:affine}
\beta_i(\qt) = p_{i,\min} + \frac{p_{i,\max} - p_{i,\min}}{q_{i,\max} - q_{i,\min}}\,(q_{t,i} - q_{i,\min}), \quad i = 1, 2,
\end{equation}
which linearly maps the spatial bounding box to the angular bounding box.
The geometric significance of this construction is explained in Section~\ref{sec:coverage}.

\subsection{Ray--reflector intersection and coverage condition}\label{sec:coverage}

Consider a ray emitted from $(\qs, 0)$ in unit direction $\shat = G^{-1}(\ps)$, described by the ray equation
\begin{equation}\label{eq:ray}
\mathbf{x}(\tau) = (\qs, 0) + \tau\, \shat, \qquad \tau \geq 0.
\end{equation}
The ray hits the reflector at parameter point $\qt$ if $\mathbf{x}(\tau) = \mathbf{r}(\qt)$ for some $\tau \geq 0$.
The third component of this equality determines the ray parameter, $\tau = u(\qt)\, d_3(\qt)/s_3$; substituting $\tau$ into the first two components and using $(s_1, s_2) = s_3\, \ps$ from~\eqref{eq:gnomonic} gives
\begin{equation}\label{eq:shadow}
\qs = \qt + u(\qt)\, d_3(\qt)\, (\boldsymbol{\beta}(\qt) - \ps).
\end{equation}
This is the key geometric relation.
It is \textit{affine} in $\ps$ for fixed $\qt$; this is the reason for working in gnomonic coordinates, since in stereographic or other nonlinear angular coordinates the corresponding equation would be nonlinear in the angular variable.
We define the \textbf{shadow map} $F_{\ps}(\qt) = \qt + u(\qt)\, d_3(\qt)\, (\boldsymbol{\beta}(\qt) - \ps)$, which gives the source-plane footprint of a reflector point $\qt$ as seen from emission direction~$\ps$.

For the degenerate profile $u \equiv 0$ the shadow map reduces to the identity, $F_{\ps} = \mathrm{id}$, which trivially covers every source point, regardless of~$\ps$.
As the profile grows continuously from zero to $u$, the shadow map keeps covering all of $\Omega$ as long as the image of the boundary $\partial\Omega$ never folds inward across $\Omega$.
This requirement yields the \textbf{outward-lean boundary condition}: for every $\mathbf{q}_b \in \partial\Omega$ with outward unit normal $\mathbf{n}$,
\begin{equation}\label{eq:boundary}
\mathbf{n} \cdot \boldsymbol{\beta}(\mathbf{q}_b) \geq \sup_{\ps \in A} \mathbf{n} \cdot \ps.
\end{equation}
The interpretation is that the reflector direction at each boundary point must lean at least as far outward (in gnomonic coordinates) as the most oblique incoming ray in that normal direction.
When this condition holds, every emitted ray---from any source point $\qs \in \Omega$ in any direction $\ps \in A$---is guaranteed to hit the reflector.
On rectangular domains, the affine map~\eqref{eq:affine} meets the condition exactly.
If $\Omega$ or $A$ is not rectangular (e.g., a disk), we simply apply the affine map to their axis-aligned bounding boxes; the direction field then leans somewhat farther outward than strictly necessary, but coverage is guaranteed, which is what we actually care about for the sake of optimization.

\subsection{Forward ray tracing and the design problem}\label{sec:design}

Given a source sample $(\qs, \ps) \in \mathcal{S}$, the forward ray-tracing pipeline proceeds as follows.
\begin{enumerate}[nosep]
\item \textbf{Emit:} ray origin $(\qs, 0)$, direction $\shat = G^{-1}(\ps)$.
\item \textbf{Intersect:} find the intersection of the ray with the reflector surface~$\mathbf{r}$.
\item \textbf{Reflect:} $\that = \shat - 2(\shat \cdot \nhat)\, \nhat$, where $\nhat$ is the unit surface normal at the intersection point; since $|\shat| = 1$, the reflected direction $\that$ is again a unit vector.
\item \textbf{Project:} compute the output angular coordinate $\pt = P_N(\that)$ via north-pole stereographic projection (since reflected rays have $t_3 < 0$):
\end{enumerate}
\begin{equation}\label{eq:stereo}
\pt = \frac{(t_1, t_2)}{1 - t_3}.
\end{equation}
This defines the forward map $\mathbf{T}_{\thv} : \mathcal{S} \to \mathcal{T}$, $(\qs, \ps) \mapsto (\qt, \pt)$, where $\thv$ denotes the parameters of $u$ (specified in Section~\ref{sec:mlp}), $B \subset \R^2$ is the far-field angular target domain (in north-pole stereographic coordinates), and $\mathcal{T} := \Omega \times B$ is the full target domain.

The map $\mathbf{T}_{\thv}$ pushes the source distribution $f$ on $\mathcal{S}$ forward to a joint distribution $\tilde{g}_{\thv}(\qt, \pt)$ on~$\mathcal{T}$.
The physically relevant quantity is the \textit{marginal} far-field angular distribution $g_{\thv}(\pt)$, obtained by integrating out the spatial reflector coordinate~$\qt$.
Via the change-of-variables formula applied to $\mathbf{T}_{\thv}^{-1}$ (cf.~\cite[Eq.~(1)]{Hacking42} and~\cite[Eq.~(7)]{Hacking41}),
\begin{equation}\label{eq:cov}
g_{\thv}(\pt) = \int_\Omega f\bigl(\mathbf{T}_{\thv}^{-1}(\qt, \pt)\bigr) \left|\det \frac{\partial \mathbf{T}_{\thv}^{-1}}{\partial (\qt, \pt)}\right| \,\dd\qt.
\end{equation}
This $g_{\thv}(\pt)$ represents the far-field angular intensity distribution produced by the reflector with profile~$u_{\thv}$.
The design problem is to find $u$ (parameterized by~$\thv$) such that $g_{\thv}(\pt) \approx \hat{g}(\pt)$ for a prescribed target distribution~$\hat{g}$.
Evaluating~\eqref{eq:cov} directly requires the inverse map $\mathbf{T}_{\thv}^{-1}$ and a two-dimensional quadrature over $\Omega$ for each target point $\pt$; Section~\ref{sec:comparison} discusses this approach and its limitations.
Instead, we adopt a \textbf{sample-based forward approach}: the source phase space is covered by $N$ fixed quasi-Monte Carlo samples, each carrying the local source density~$f$ as a weight, and these are ray-traced through the reflector via~$\mathbf{T}_{\thv}$; the resulting weighted output distribution is compared to $\hat{g}$ via a differentiable loss (Section~\ref{sec:histogram}).

\begin{figure}[htbp]
\centering
\includegraphics[width=0.45\linewidth]{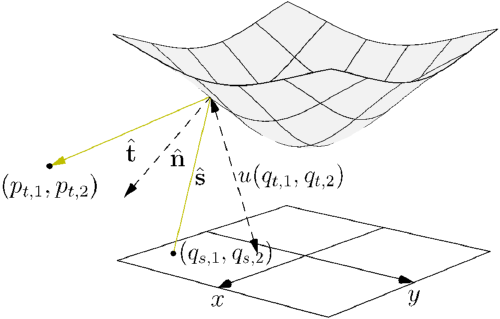}
\caption{Schematic diagram of the 3D finite-source reflector problem. A planar source on $z = 0$ emits rays from spatial points $\qs = (q_{s,1}, q_{s,2})$ in directions $\shat$. Each ray strikes the reflector surface $\mathbf{r}(\qt) = (\qt, 0) + u(\qt)\,\dhat(\qt)$, reflects about the surface normal $\nhat$, and propagates to the far field. The reflected direction $\that$ is projected stereographically to give the far-field angular coordinate $\pt = (p_{t,1}, p_{t,2})$.}
\label{fig:system_diagram}
\end{figure}

\begin{table}[htbp]
\caption{Notation and symbols.}
\label{tab:notation}
\centering
\small
\begin{tabular}{@{} >{$}l<{$} p{0.62\textwidth} @{}}
\toprule
\textrm{Symbol} & Meaning \\
\midrule
\Omega & Spatial support of the planar source (source base domain) \\
A & Angular source domain (gnomonic coordinates) \\
\mathcal{S} := \Omega \times A & Full source domain \\
B & Far-field angular target domain (north-pole stereographic coordinates) \\
\mathcal{T} := \Omega \times B & Full target domain \\
f(\qs, \ps) & Normalized source distribution function over $\mathcal{S}$ \\
\thv & Parameters of the profile function $u$ (e.g., MLP weights and biases) \\
\tilde{g}_{\thv}(\qt, \pt) & Pushforward of $f$ under $\mathbf{T}_{\thv}$; joint distribution over $\mathcal{T}$ \\
g_{\thv}(\pt) & Marginal far-field angular distribution (integral of $\tilde{g}_{\thv}$ over $\qt$) \\
\hat{g}(\pt) & Prescribed (desired) far-field angular target distribution \\
\qs & Source spatial coordinate in $\Omega$ \\
\qt & Reflector surface parameter in $\Omega$ \\
\ps & Gnomonic emission coordinate in $A$ \\
\pt & Stereographic far-field coordinate in $B$ \\
\shat & Unit emission direction ($\in \Sph^2_+$) \\
\that & Unit reflected direction ($t_3 < 0$) \\
\mathbf{r}(\qt) & Reflector surface point for parameter $\qt$ \\
u(\qt) & Reflector profile function \\
\dhat(\qt) & Unit direction field of the reflector ($\in \Sph^2_+$) \\
\boldsymbol{\beta}(\qt) & Gnomonic representation of the direction field \\
\nhat & Unit normal to the reflector surface \\
\mathbf{T} & Forward map from source to target, $\mathcal{S} \to \mathcal{T}$; written $\mathbf{T}_{\thv}$ when emphasizing dependence on the profile parameters \\
G,\; G^{-1} & Gnomonic projection and its inverse \\
P_N & North-pole stereographic projection \\
\bottomrule
\end{tabular}
\end{table}

\section{Method}\label{sec:method}

This section details the four components of our approach: the neural-network parameterization of the reflector profile (Section~\ref{sec:mlp}), the spline-compiled surface and GPU ray tracing (Section~\ref{sec:optix}), the $H^{-1}$-weighted soft-histogram loss (Section~\ref{sec:histogram}), and the quasi-Newton optimizer (Section~\ref{sec:bfgs}).
Section~\ref{sec:comparison} discusses the relationship to prior lower-dimensional formulations.

\subsection{Neural-network parameterization of the profile function}\label{sec:mlp}

We parameterize the profile function $u$ by an MLP with parameters~$\thv$.
The network takes as input the two-dimensional reflector coordinate $(q_{t,1}, q_{t,2})$, passes it through two hidden layers of 24 units each with the activation function $\phi(x) := \tanh^2(x)$, and produces a single scalar output.
The raw network output is combined with learnable scale, quadratic, and bias terms and passed through a sigmoid function to ensure $u \in [u_{\min}, u_{\max}]$:
\begin{equation}\label{eq:mlp}
u_{\thv}(\qt) = u_{\min} + (u_{\max} - u_{\min})\, \sigma\!\bigl(\lambda\, \mathrm{MLP}_{\thv}(\qt) + \gamma\, |\qt|^2 + h_{\mathrm{logit}}\bigr),
\end{equation}
where $\sigma$ is the sigmoid function, $\lambda$ (initialized to~0.1) is a learnable linear scale, $\gamma$ (initialized to~0.5) is a learnable quadratic scale, and $h_{\mathrm{logit}}$ is initialized so that $\sigma(h_{\mathrm{logit}}) = (u_{\mathrm{init}} - u_{\min})/(u_{\max} - u_{\min})$.
Together with the network's 697 weights and biases, $\thv$ comprises 700 trainable parameters.
In all experiments we set $u_{\min} = 0.1$, $u_{\max} = 8.0$, and $u_{\mathrm{init}} = 2.0$; the same initial profile value is used for every example.

At initialization, $\mathrm{MLP}_{\thv}(\qt) \approx 0$ across $\Omega$ and $\lambda$ is small, so $u_{\thv}(\qt) \approx u_{\mathrm{init}}$ up to a small quadratic perturbation.
This yields a near-constant profile that places the reflector at approximately uniform distance from the source plane along the direction field, providing a physically reasonable starting point for the optimization.

\subsection{Spline-compiled surface and GPU ray tracing}\label{sec:optix}

Instead of triangulating the surface, the MLP profile $u_{\thv}$ is \textit{compiled} each iteration: it is evaluated on a regular $64 \times 64$ grid over the padded bounding box of~$\Omega$, and the resulting values define a bicubic B-spline interpolant of the profile.
All ray tracing is performed against this smooth approximation surface directly.
Because the intersection relation~\eqref{eq:shadow} is affine in $\ps$ and only slightly nonlinear in $\qt$, the intersection parameter $\qt$ of each ray is found by a damped two-dimensional Newton iteration.
This requires relatively few iterations to converge to a solution that is sufficiently accurate for our applications.
The surface normal at the hit point then comes directly from the analytic gradient of the spline interpolant---though it is of course only an approximation of the MLP's normal at that point.

Gradient flow through the intersection uses the implicit function theorem: differentiating~\eqref{eq:shadow} at the converged point yields the sensitivity of $\qt$ (and thus the reflected direction) to local perturbations of the reflector profile and its normal in closed form.
Simply using automatic differentiation here---which would differentiate through the iterative algorithm---might still be sufficient as well, but the implicit function approach should be faster and more accurate, especially as the number of inner Newton iterations is increased.
The complete pipeline---reflector profile compilation, Newton intersection, reflection, histogram accumulation---runs on a single GPU.
On our hardware, a full loss-and-gradient evaluation with $N = 2^{22}$ rays takes well under a millisecond.

\subsection{$H^{-1}$-weighted soft-histogram loss}\label{sec:histogram}

We require a differentiable loss that compares the weighted empirical distribution of the $N$ ray-traced output samples $\{\pt^{(i)}\}_{i=1}^{N}$ against a target distribution~$\hat{g}$.
Standard histograms with hard indicator-function bins are non-differentiable at the bin boundaries.
We therefore accumulate rays in a `soft' histogram: each sample deposits its source weight onto a grid of $M \times M$ bins ($M = 128$) through compactly supported B-spline kernels, such that the bin contents $h_{jk}(\thv)$ are smooth functions of the sample positions.
This is necessary in order for the objective function to be optimized through gradient-based optimization algorithms.
We choose this B-spline kernel as it is both smooth and compactly supported, which keeps the computation relatively cheap and comparable to a standard histogram.
Related differentiable histogram constructions have been explored by Avi-Aharon et al.~\cite{AviAharonArbelleRaviv36}, and B-spline-based density estimation by Kirkby et al.~\cite{KirkbyLeitaoNguyen37} and Zhao et al.~\cite{ZhaoZhangNiWang38}.
A reference histogram $h^{\mathrm{ref}}$ is precomputed once from the target distribution using the same binning and B-spline kernel, such that both sides of the comparison carry the same kernel smoothing.
This ensures that an exact solution---if possible---would obtain a discrepancy of exactly $0$.

Rather than penalizing the residual $h - h^{\mathrm{ref}}$ uniformly across frequencies, we measure it under an $H^{-1}$-type spectral weighting (a weighting that emphasizes low spatial frequencies),
\begin{equation}\label{eq:loss}
\mathcal{L}(\thv) = \frac{1}{\|h^{\mathrm{ref}}\|_2^2\,M^2} \sum_{\mathbf{k}} \left(\frac{1}{1 + |\mathbf{k}|^2/k_0^2} + \epsilon_0\right) \bigl|\widehat{(h - h^{\mathrm{ref}})}(\mathbf{k})\bigr|^2,
\end{equation}
where $\widehat{\,\cdot\,}$ denotes the discrete Fourier transform over the bin grid, $\mathbf{k}$ the spatial frequency, $k_0 = 1$ (one cycle across the histogram window), and $\epsilon_0 = 0.02$ a small flat floor.
The low-pass factor is the discrete analogue of a squared $H^{-1}$ norm and encodes the transport nature of the problem, without requiring expensive Wasserstein distance or Sinkhorn divergence computations to be performed repeatedly.
This formulation ensures that, even when the achieved and desired distributions are far apart in terms of Wasserstein distance, there is still a good gradient signal pushing the ray-traced distribution in the right direction by allowing it to consider longer-range transport, which more standard losses might miss.
The final auxiliary loss term simply encourages the optimizer to push as much reflected flux as possible into the desired domain, which the main histogram loss is blind to.
Each ray that fails to reach the desired target domain contributes to an escaping-flux fraction, with a weight that grows smoothly from zero with the distance by which that ray overshoots the domain boundary.
This squared fraction is then added to the loss.

For evaluation purposes (as opposed to training), we use a standard hard histogram with $M_{\mathrm{eval}} = 128$ bins per axis and report the \textbf{relative $L^2$ error}
\begin{equation}\label{eq:l2error}
\varepsilon_{L2} = \frac{\|h - h^{\mathrm{ref}}\|_2}{\|h^{\mathrm{ref}}\|_2},
\end{equation}
where $h$ and $h^{\mathrm{ref}}$ now denote the hard histograms at resolution $M_{\mathrm{eval}}$, computed from $2^{26}$ fresh quasi-Monte Carlo samples independent of those used during training.

\subsection{Quasi-Newton optimization}\label{sec:bfgs}

The quasi-Newton update uses a self-scaled Broyden family method.
Self-scaled quasi-Newton methods, introduced by Oren and Luenberger~\cite{OrenLuenberger35} and further developed by Al-Baali~\cite{AlBaali39}, multiply the current Hessian-inverse approximation by a scalar factor before performing the standard secant update.
We use the self-scaled Broyden variant of Urb\'an et al.~\cite{UrbanStefanouPons30}, which performed best in their study of quasi-Newton training for physics-informed neural networks, where such methods substantially outperformed first-order optimizers such as Adam~\cite{KingmaBa40}.
Step sizes come from the bracketing weak-Wolfe line search of Lewis and Overton~\cite{LewisOverton43}.
The initial step size is warm-started from the previously accepted line-search step size, and the gradients of line-search trial points are computed only when the sufficient-decrease (Armijo) condition is satisfied to avoid unnecessary gradient evaluations for rejected trials.
When the search fails, which happens on non-convex landscapes, the optimizer falls back to a Hessian reset, then Armijo backtracking, then a fixed step.

Two mechanisms improve the reliability of the optimization on this non-convex landscape.
First, a \textit{plateau-perturbation} (basin-hopping) rule: whenever progress stalls, the iterate is perturbed by a small random step (with amplitude decreasing over successive perturbations, on a deterministic schedule) and the Hessian approximation is reset; each perturbation launches from---and the run finally returns to---the best iterate encountered, so the best solution found is never lost.
This reliably restores progress where plain descent otherwise stalls.
Runs initialized from the solution at a previous radius (the continuation strategy of Section~\ref{sec:example_c}) skip these perturbations: the continuation exists to keep the solution in the inherited basin, and perturbing a near-converged solution tends to find basins that improve the training loss only by exploiting the particular fixed sample set.
Second, we penalize large pre-sigmoid values in~\eqref{eq:mlp}: a profile pushed against $u_{\min}$ or $u_{\max}$ saturates the sigmoid, so gradients through it vanish and the optimizer cannot pull the profile back from the bound.
Because the penalty acts on the pre-sigmoid values directly, it keeps pushing the profile away from the bounds even where the sigmoid has saturated.

Source samples are generated once from a deterministic low-discrepancy point set (a rank-one lattice with a seed-dependent random shift) and held fixed throughout optimization.
This deterministic sampling is necessary for standard quasi-Newton methods, as they assume deterministic objectives; stochastic objectives would corrupt the curvature information and break the assumptions of the line search.
Stochastic quasi-Newton variants~\cite{MoritzNishiharaJordan31} do exist, but we opted here for determinism as the ray tracer is sufficiently fast to allow for large sample sizes, and the deterministic objective then allows us to converge faster and more reliably.

\subsection{Relation to lower-dimensional formulations}\label{sec:comparison}

Earlier lower-dimensional formulations~\cite{Hacking41} can---to some extent---exploit symmetry to reduce the source and target phase spaces from four dimensions (2D spatial $\times$ 2D angular) to two dimensions (1D spatial $\times$ 1D angular), which enables inverse-mapping- and quadrature-based losses at manageable cost.
In the fully three-dimensional setting treated here, those constructions get expensive: the marginal far-field distribution in~\eqref{eq:cov} needs a two-dimensional integral over~$\Omega$ for each target location, and constructions based on intersecting source and target cells would have to operate in the four-dimensional phase space.
Such constructions may still be extended to 3D, but would require more careful attention to the computational cost and efficiency of the quadrature algorithms to avoid excessive runtime.

The present method therefore takes a different route: a forward, sample-based pipeline with hardware-accelerated ray tracing and a smooth soft-histogram discrepancy.
It evaluates no inverse maps and no four-dimensional cell intersections, has no trouble with discontinuous source distributions, and suits GPU-based quasi-Newton optimization.
Both lines of work use compact neural parameterizations and quasi-Newton updates, but the objective construction and the optimization loop have little else in common.

\section{Numerical results}\label{sec:results}

We demonstrate the method on three examples of increasing complexity: a Gaussian-to-uniform redistribution (Section~\ref{sec:example_a}), a yin-yang image projection (Section~\ref{sec:example_b}), and a systematic sweep over the angular source extent (Section~\ref{sec:example_c}).
Implementation details common to all examples are given in Section~\ref{sec:implementation}.

\subsection{Implementation details}\label{sec:implementation}

The method is implemented in CUDA C++: profile compilation, Newton intersection, reflection, histogram accumulation, the adjoint, and the network itself all execute on the GPU, with the optimizer state held in double precision on the host.
Surface and ray computations use single precision (float32) on a single NVIDIA RTX~4090 GPU\@.
All examples use $N = 2^{22}$ quasi-Monte Carlo source samples and $128 \times 128$ soft-histogram bins, and there are no per-example settings: every hyperparameter, including the initial profile value and the loss-histogram resolution, is identical across the three examples.
The only per-example difference is the wall-clock budget, which serves as the stopping criterion: 2 seconds per run for Examples~A and~B, and 8 seconds per radius for Example~C.
Because the order of parallel floating-point additions varies from run to run on a GPU, histogram and adjoint sums are accumulated in fixed-point integer arithmetic (integer addition is exactly associative), so the objective---and with it the entire optimization trajectory---is bitwise reproducible across runs on a given GPU.

\subsection{Example A: truncated Gaussian source, uniform target}\label{sec:example_a}

The first example uses a square spatial domain $\Omega = [-0.3, 0.3]^2$ and a circular angular domain $A$ of gnomonic radius~$0.75$ (cone half-angle $\arctan 0.75 \approx 37^\circ$).
The source distribution is a truncated Gaussian with zero mean and covariance matrix $0.2\, I_4$ on $\Omega \times A$, and the target is a uniform distribution $\hat{g}$ on $[-0.3, 0.3]^2$ in north-pole stereographic coordinates.

\begin{figure}[p]
\centering
\makebox[\textwidth][c]{\includegraphics[width=1.13\textwidth]{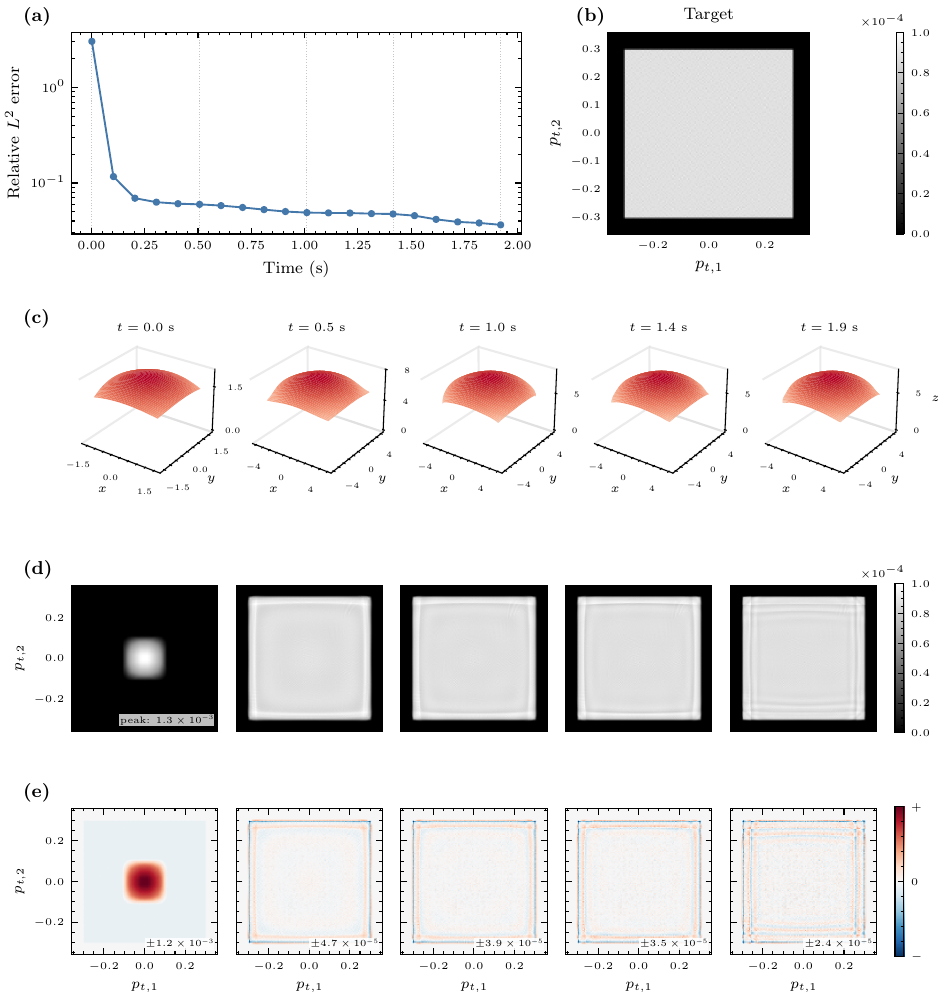}}
\caption{Example~A: truncated-Gaussian source, uniform target. (a)~Relative $L^2$ error over wall-clock time; the vertical dotted lines mark the snapshot times shown below. (b)~Target distribution. (c)~Reflector surface at selected times. (d)~Ray-traced far-field distributions; the first panel uses its own color scale (peak annotated), while the remaining panels share a common scale with~(b). (e)~Signed error $e = h - h^{\mathrm{ref}}$, individually scaled to $\pm\max|e|$ per panel.}
\label{fig:example_a}
\end{figure}

The optimizer reaches a relative $L^2$ error of $\varepsilon_{L2} = 3.6\%$ within the 2-second budget.
Figure~\ref{fig:example_a} shows the convergence history, the evolution of the reflector surface and the ray-traced far-field distribution at selected times, and the pointwise signed error with respect to the target.
The error drops below $7\%$ within the first quarter second and continues to improve steadily until the budget is exhausted.
Throughout the optimization, the optimizer adapts the curvature of the reflector to redistribute flux from the center---where the Gaussian source is densest---toward the edges, in order to match the prescribed uniform target.

\subsection{Example B: uniform source, yin-yang image target}\label{sec:example_b}

The second example uses a circular spatial domain $\Omega$ of radius~$0.3$ and the same circular angular domain $A$ of gnomonic radius~$0.75$ as Example~A, with a uniform source distribution on $\Omega \times A$.
The target distribution $\hat{g}$ is a yin-yang image ($512 \times 512$ grayscale, piecewise-constant interpolation) defined on $[-0.3, 0.3]^2$.

\begin{figure}[p]
\centering
\makebox[\textwidth][c]{\includegraphics[width=1.13\textwidth]{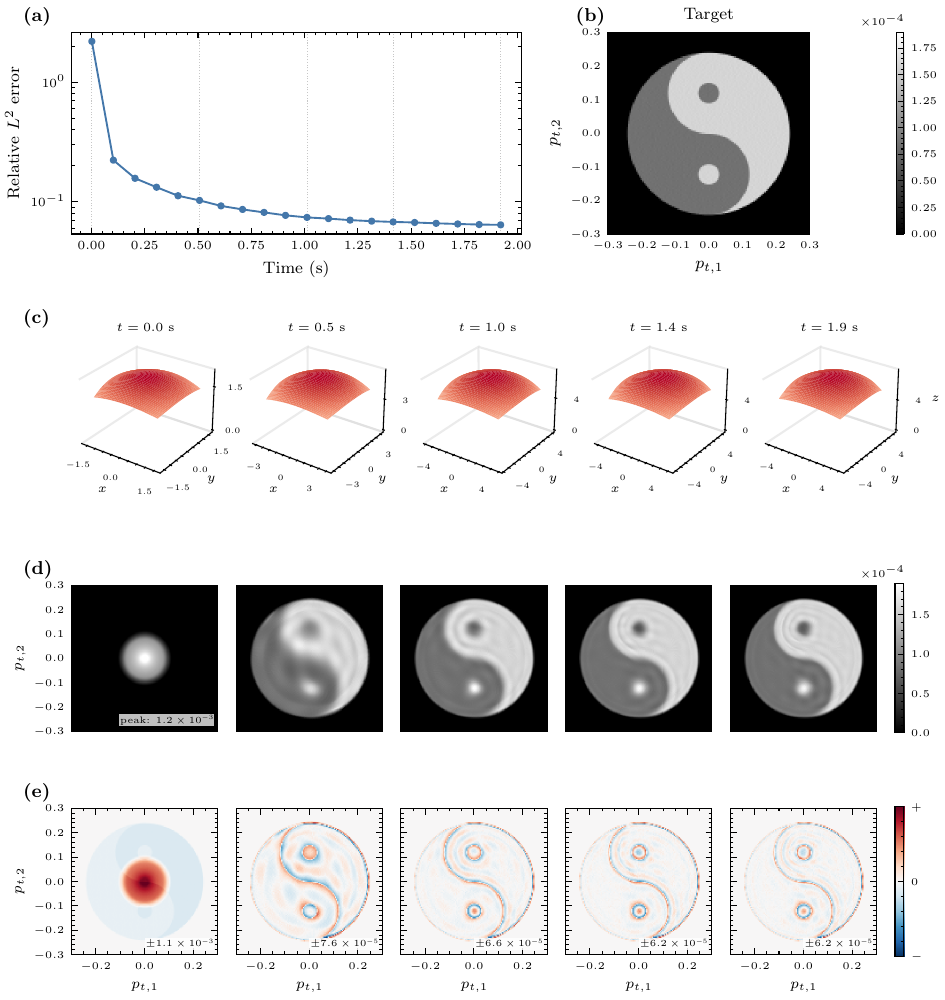}}
\caption{Example~B: uniform source, yin-yang image target. (a)~Relative $L^2$ error over wall-clock time; the vertical dotted lines mark the snapshot times shown below. (b)~Target distribution. (c)~Reflector surface at selected times. (d)~Ray-traced far-field distributions; the first panel uses its own color scale (peak annotated), while the remaining panels share a common scale with~(b). (e)~Signed error, individually scaled to $\pm\max|e|$ per panel.}
\label{fig:example_b}
\end{figure}

The optimizer converges to $\varepsilon_{L2} = 6.4\%$ within the 2-second budget.
The results are shown in Figure~\ref{fig:example_b}.
The yin-yang pattern emerges progressively from the initial near-uniform distribution as the reflector surface is shaped by the optimizer.
The final result captures the main features of the prescribed yin-yang target, namely the two interlocking regions and the small dots contained within, though with visible blurring and some ringing artifacts, which we believe are inevitable given the non-zero \'etendue of the source. As an exact match is likely impossible, the optimizer instead finds a solution that minimizes the specified objective. Alternative objectives may therefore yield different trade-offs between the various features of the target, though we leave such exploration to future work.

\subsection{Example C: angular domain radius sweep}\label{sec:example_c}

Examples~A and~B both use a fixed angular domain radius of $r_A = 0.75$.
To study how the optimization result depends on the angular source extent, Example~C systematically varies this angular radius over a wide range.
The spatial domain is a disk $\Omega$ of radius~$0.3$, the source distribution is uniform on $\Omega \times A(r_A)$ where $A(r_A)$ denotes a disk of gnomonic radius~$r_A$, and the target is uniform on $[-0.3, 0.3]^2$.
The radius is swept over ten equally spaced values $r_A = 2k/9$, $k = 0, \dots, 9$, from the collimated limit $r_A = 0$ up to $r_A = 2$ (cone half-angle $\approx 63^\circ$).
Apart from the collimated case, which is solved independently, we proceed by continuation: we first optimize at the smallest non-zero radius, then sweep outward, initializing each solve from the solution at the preceding radius.

\begin{figure}[p]
\centering
\makebox[\textwidth][c]{\includegraphics[width=1.13\textwidth]{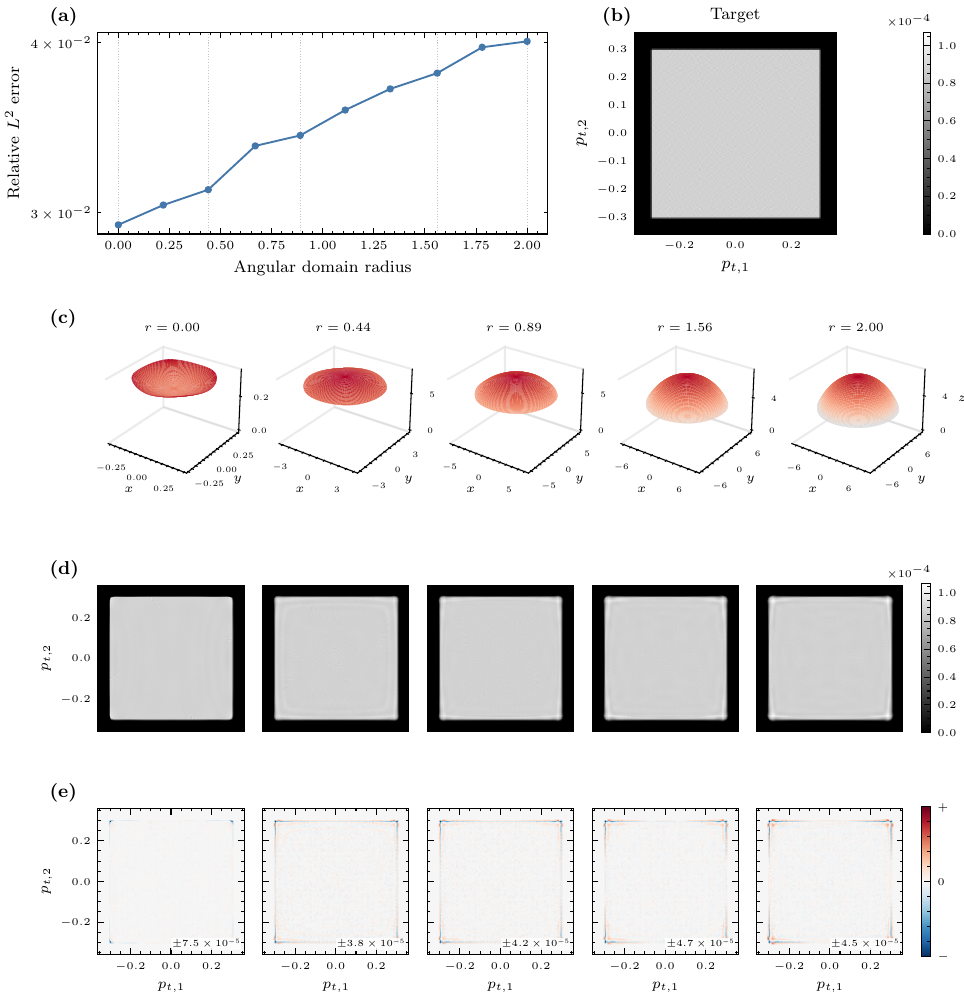}}
\caption{Example~C: angular domain radius sweep. (a)~Relative $L^2$ error versus angular radius; the vertical dotted lines mark the radii shown below. (b)~Target distribution. (c)~Reflector surface at selected radii, sampled over the illuminated source disk; each panel uses its own $z$-axis scale (annotated). (d)~Ray-traced far-field distributions (common color scale). (e)~Signed error, individually scaled to $\pm\max|e|$ per panel.}
\label{fig:example_c}
\end{figure}

The results are shown in Figure~\ref{fig:example_c} and summarized in Table~\ref{tab:results}.
At $r_A = 0$ (a parallel source in which all rays travel in the same direction), the relative $L^2$ error is $\varepsilon_{L2} = 2.9\%$, the smallest across the sweep, consistent with the fact that the parallel case approaches a collimated-beam reflector problem for which highly accurate solutions are known to exist~\cite{BoselGross11,BoselWorkuGross12}.
As $r_A$ increases, the error grows \textit{monotonically} and gently, from $3.0\%$ at $r_A = 0.22$ to $4.0\%$ at $r_A = 2.0$---a factor of only $1.3$ while the angular source extent grows by an order of magnitude.
The monotone growth of the error is consistent with our physical expectation: as the angular extent of the source is increased, the attainable sharpness of the target is limited by the source \'etendue, and the optimizer must find solutions that are near this limit across the entire sweep.

\begin{table}[htbp]
\caption{Summary of numerical results.}
\label{tab:results}
\centering
\footnotesize
\setlength{\tabcolsep}{4pt}
\begin{tabular}{@{} l l l l l r r r @{}}
\toprule
Example & $\Omega$ & $A$ & Source & Target & $N$ & $\varepsilon_{L2}$ & Time (s) \\
\midrule
A & box $[-0.3,0.3]^2$ & disk, $r\!=\!0.75$ & Trunc.\ Gauss. & Uniform & $2^{22}$ & 3.6\% & 2.0 \\
B & disk, $r\!=\!0.3$ & disk, $r\!=\!0.75$ & Uniform & Yin-yang & $2^{22}$ & 6.4\% & 2.0 \\
C ($r_A \!=\! 0$) & disk, $r\!=\!0.3$ & disk, $r_A \!=\! 0$ & Uniform & Uniform & $2^{22}$ & 2.9\% & 8.0 \\
C ($r_A \!=\! 2$) & disk, $r\!=\!0.3$ & disk, $r_A\!=\!2$ & Uniform & Uniform & $2^{22}$ & 4.0\% & 8.0 \\
\bottomrule
\end{tabular}
\end{table}

\section{Discussion}\label{sec:discussion}

In the numerical experiments the method converges within 2 to 8 seconds of wall-clock time on a single GPU.
Three things make this possible: the spline-compiled surface with its cheap Newton intersection (a full loss-and-gradient evaluation with $2^{22}$ rays takes well under a millisecond), the fully GPU-resident pipeline, and the compactly supported soft-histogram accumulation.

While those three components ensure that objective and gradient computations are fast, it is also important that our optimization can converge reliably in relatively few iterations.
We believe there are three main components that allow for this.
Firstly, our use of gnomonic coordinates yields an affine ray--reflector intersection equation, which ensures all rays intersect the reflector regardless of the current neural network parameter vector, and avoids any potential discontinuities in the objective function resulting from varying visibility.
Secondly, the $H^{-1}$-weighted loss additionally improves the reliability of the optimization by supplying a long-range transport signal, which dominates during early optimization (when there is a greater mismatch between achieved and desired target distributions).
Finally, our use of plateau-perturbation helps to eliminate any further failures or premature optimizer terminations which we otherwise sometimes observed.

Compared with prior lower-dimensional inverse-mapping formulations, the present approach uses a forward ray-tracing pipeline combined with a soft-histogram loss.
This change is motivated by the increased dimensionality: in 3D without rotational symmetry, evaluating the marginal far-field distribution $g_{\thv}(\pt)$ via~\eqref{eq:cov} requires repeated two-dimensional quadrature over~$\Omega$ for each far-field point, and the integrand develops singular behavior as the angular source domain shrinks.
The forward approach avoids these issues entirely and, being based on sample statistics rather than pointwise density evaluation, naturally handles discontinuous source distributions.
Compact MLP parameterizations (approximately 700 parameters) remain effective in this fully three-dimensional setting.
As noted previously, efficient inverse-mapping constructions may still be effective, but would require carefully designed domain-specific quadrature methods to reduce the number of evaluation points if we wish to achieve performance similar to or better than that of the present forward map approach.

Several limitations of the current method should be noted.
The optimization landscape is non-convex, and the quality of the reached optimum can still depend on the basin entered early in the run; the bitwise-reproducible pipeline implemented here makes this dependence deterministic rather than stochastic (at least within the same environment), but it surfaces as sensitivity to the initial profile height: in Example~A, different choices of $u_{\mathrm{init}}$ reach local optima with errors between $3.2\%$ and $6.5\%$.
In the collimated limit of Example~C, the residual error concentrates at the corners of the target, where the ray mapping must carry the smooth circular boundary of the source disk into the sharp corners of the square.
The smooth neural network struggles to represent such examples accurately, though for larger \'etendue cases this is less relevant, as it is no longer the dominant error.
Finally, the profile is represented through a fixed $64 \times 64$ spline compilation grid, which bounds the finest surface detail the optimizer can express.

We see several directions for future work.
An adaptive spline compilation grid, concentrating resolution where the reflector curvature is highest, could improve accuracy at little extra cost, though at increased implementational complexity.
Convexity-structured parameterizations, in the spirit of the convexity-constrained neural {Monge--Amp\`ere} solver of~\cite{Hacking29}, could reduce the remaining basin dependence of the non-convex landscape.
Additionally, the currently unconstrained reflector shape may develop geometries that result in rays bouncing multiple times within the reflector, which the current model does not account for.
Though we have not observed this problem during our experiments, constrained geometries (such as convexity constraints) may still be useful for more formally ensuring that rays may not bounce multiple times within the reflector.
Finally, as noted previously, non-zero \'etendue sources may no longer admit exact solutions, and our choice of objective function therefore influences how we distribute this irreducible error.
For example, we may wish to reduce the ringing observed in some of our results, and future research might therefore investigate alternative distribution discrepancies to distribute the error in a more desirable fashion.

\section{Conclusion}\label{sec:conclusion}

We have presented a direct optimization method for fully three-dimensional finite-source reflector design, built on a neural surface parameterization compiled to smooth spline surfaces.
The efficient differentiable ray-tracing formulation we developed around this representation, based on an $H^{-1}$-weighted soft-histogram loss, allows us to efficiently optimize close-to-zero \'etendue sources, as well as sources with both significant spatial and angular extent.
Our method converges within just a few seconds on a single GPU across several examples, using a deterministic and exactly reproducible pipeline.


\begin{backmatter}
\bmsection{Funding}
This work in the project MALIOD is funded by Holland High Tech---TKI HSTM via the PPS allowance scheme for public--private partnerships.

\bmsection{Disclosures}
The authors declare no conflicts of interest.

\bmsection{Data availability}
Data underlying the results presented in this paper are not publicly available at this time but may be obtained from the authors upon reasonable request.
\end{backmatter}

\end{document}